\documentclass[useAMS,usenatbib]{mn2e}
\usepackage{natbib}
\usepackage{amsmath,amsfonts}
\usepackage{epsfig}
\def\reff@jnl#1{{\rm#1\/}}

\def\aj{\reff@jnl{AJ}}                  % Astronomical Journal
\def\araa{\reff@jnl{ARA\&A}}            % Annual Review of Astron and Astrophys
\def\apj{\reff@jnl{ApJ}}                        % Astrophysical Journal
\def\apjl{\reff@jnl{ApJ}}               % Astrophysical Journal, Letters
\def\apjs{\reff@jnl{ApJS}}              % Astrophysical Journal, Supplement
\def\ao{\reff@jnl{Appl.Optics}}         % Applied Optics
\def\apss{\reff@jnl{Ap\&SS}}            % Astrophysics and Space Science
\def\aap{\reff@jnl{A\&A}}               % Astronomy and Astrophysics
\def\aapr{\reff@jnl{A\&A~Rev.}}         % Astronomy and Astrophysics Reviews
\def\aaps{\reff@jnl{A\&AS}}             % Astronomy and Astrophysics, Supplement
\def\azh{\reff@jnl{AZh}}                        % Astronomicheskii Zhurnal
\def\baas{\reff@jnl{BAAS}}              % Bulletin of the AAS
\def\jrasc{\reff@jnl{JRASC}}            % Journal of the RAS of Canada
\def\memras{\reff@jnl{MmRAS}}           % Memoirs of the RAS
\def\mnras{\reff@jnl{MNRAS}}            % Monthly Notices of the RAS
\def\pra{\reff@jnl{Phys. Rev. A}}         % Physical Review A: General Physics
\def\prb{\reff@jnl{Phys. Rev. B}}         % Physical Review B: Solid State
\def\prc{\reff@jnl{Phys. Rev. C}}         % Physical Review C
\def\prd{\reff@jnl{Phys. Rev. D}}         % Physical Review D
\def\prl{\reff@jnl{Phys. Rev. Lett}}      % Physical Review Letter
\def\pasp{\reff@jnl{PASP}}              % Publications of the ASP
\def\pasj{\reff@jnl{PASJ}}              % Publications of the ASJ
\def\qjras{\reff@jnl{QJRAS}}            % Quarterly Journal of the RAS
\def\skytel{\reff@jnl{S\&T}}            % Sky and Telescope
\def\solphys{\reff@jnl{Solar~Phys.}}    % Solar Physics
\def\sovast{\reff@jnl{Soviet~Ast.}}     % Soviet Astronomy
\def\ssr{\reff@jnl{Space~Sci.Rev.}}     % Space Science Reviews
\def\zap{\reff@jnl{ZAp}}                        % Zeitschrift fuer Astrophysik
\def\nat{\reff@jnl{Nature}}             % Nature 

\def\p#1by#2{{\partial{#1} \over \partial{#2}}}
\def\pp#1by#2#3{{\partial^2{#1} \over \partial{#2}\partial{#3}}}
\def\d#1by#2{{{\rm d}{#1} \over {\rm d}{#2}}}
\def\dd#1by#2#3{{{\rm d}^2{#1} \over {\rm d}{#2}{\rm d}{#3}}}

\title[An excess of emission in the dark cloud LDN\,1111]
{An excess of emission in the dark cloud LDN\,1111 with the Arcminute Microkelvin Imager\thanks{We request that any reference to this paper cites ``AMI Consortium: Scaife et~al. 2008"}}

\label{firstpage}
\author[Scaife et~al.]
{AMI CONSORTIUM:
 Scaife A. M. M.$\thanks{E-mail: 
as595@mrao.cam.ac.uk}$, 
 Hurley-Walker N.,
 Green D. A.,
 \newauthor
 Davies M. L., 
 Grainge K. J. B., 
 Hobson M. P.,
 Lasenby, A. N.,
 L{\'o}pez-Caniego M.,
 \newauthor
 Pooley G. G.,
 Saunders R. D. E.,
 Scott P. F.,
 Titterington D. J.,
 Waldram E M.,
\newauthor
 Zwart J. T. L.\\
 \vspace{0.03in}\\
$^1$ Astrophysics Group, Cavendish Laboratory, J J Thomson Avenue,
Cambridge CB3 0HE\\
}

\date{Accepted ---; received ---; in original form \today}
\pagerange{\pageref{firstpage}--\pageref{lastpage}}
\pubyear{2005}

\begin{document}
%\label{firstpage}
\maketitle

\begin{abstract}
\noindent
We present observations of the Lynds' dark nebula LDN\,1111 made at
microwave frequencies between 14.6 and 17.2\,GHz with the Arcminute
Microkelvin Imager (AMI). We find emission in this
frequency band in excess of a thermal free--free spectrum 
extrapolated from data at 1.4\,GHz with matched {\it
  uv}-coverage. This excess is
$> 15\sigma$ above the predicted emission. We fit the measured spectrum
using the spinning dust model of Drain \& Lazarian (1998a) and find
the best fitting model parameters agree well 
with those derived from Scuba data for this object by Visser et~al. (2001).

\end{abstract}


\section{Introduction}

Recent pointed observations (Finkbeiner et~al. 2002, 2004; Casassus et~al. 2004, 2006; Watson et al. 2005; Scaife et~al. 2007; Dickinson et~al. 2007)
have provided some evidence for the anomalous microwave emission
commonly ascribed to spinning dust (Drain \& Lazarian
1998a,b). Although this emission was originally seen as a large scale
phenomenon in CMB observations (see e.g. Kogut et~al. 1996) it has
 been suggested that the emission occurs in a number of distinct astronomical
 objects,
 such as dark clouds, and {\sc Hii} and photo-dissociation regions. It
 often appears to be
correlated with thermal dust emission as supported by the pointed
observations mentioned
 previously but it must be stated that this is not always the case, see for
 example
 Casassus et~al. (2008).

In spite of these predictions the evidence for anomalous microwave
emission in compact 
objects is often contradictory. Early observations  below 10\,GHz of
the molecular cloud LPH\,96 (Finkbeiner et~al. 2002), which showed a
rising spectrum were later contradicted by observations at 31 and
33\,GHz which found emission consistent with an optically thin
free--free spectrum extrapolated from lower frequencies 
(Dickinson et~al. 2006; Scaife et~al. 2007). Although some evidence
for an excess was found in a sample of Southern {\sc Hii} regions
(Dickinson et~al. 2007) and more significantly in RCW175 (Dickinson
et~al. 2008), no 
emission inconsistent with free--free was 
found in a sample of Northern {\sc Hii} regions (Scaife et~al. 2008).
Casassus et~al. (2004) proposed a flux density of approximately 1\,Jy from the Helix
planetary nebula at 31\,GHz to be in excess of a free--free spectrum
extrapolated from lower frequencies. However, based on flux densities
from the literature at 1.4, 2.7, 6.63\,GHz (Ehman, Dixon, 
\& Kraus 1970; Higgs 1971; Wall, Wright, 
\& Bolton 1976) and the recent WMAP 5-year
densities at 23--94\,GHz (Wright et~al. 2008), which are all also $\approx 1$\,Jy, we 
suggest that there is no evidence for this reported excess in the flux density
spectrum, although the nebula may be anomalous in other ways.

In this letter we present observations of the Lynds' dark nebula
LDN\,1111, taken from the AMI sample of compact Galactic star
formation regions (Scaife et~al., in prep). This sample was selected from
the SCUBA sample of compact Lynds' clouds (Visser, Richer \& Chandler 2001). 

The spectra of dark clouds at gigahertz frequencies is poorly
documented in all but a few cases (Finkbeiner et~al 2004; Casassus
et~al. 2006; Casassus et~al. 2008). In those cases where cm-wave data is
available (Casassus et~al. 2006; Casassus et~al. 2008) the behaviour
of these objects has been found to be anomalous in a number of ways
and in the case of LDN\,1622 (Casassus et~al. 2006) to show a distinct
excess of microwave emission.

LDN\,1111 ($\alpha = 21^{\rm{h}} 40^{\rm{m}} 30^{\rm{s}}, \delta = 
+57^{\circ} 48' 00''$ J2000) lies on the inside edge of the heel of
the large ($\approx 3$\,degree) horseshoe shaped {\sc Hii} region
IC\,1396 (Sh 2-131; Sharpless 1959). 
It has no known IRAS association (Parker 1988) but has been studied in
the submillimetre at 
850\,$\mu$m with the SCUBA instrument (Visser, Richer \&
Chandler 2001). An opacity class 6 object ($A_{\nu} \ge 5$\,mag; Lynds
1962), it is one of the most
opaque dark nebulae.

\begin{section}{Observations}

\begin{subsection}{Calibration and Data Reduction}

The AMI Small Array (SA) is a radio interferometer which observes in
eight frequency 
channels in the band 12--18\,GHz at the Mullard Radio Astronomy Observatory,
Lord's Bridge, Cambridge, UK. In practice, the lowest two frequency channels
are generally unused due to a low response in this frequency range, and
interference from geostationary satellites. \citet{0807.2469} discusses the
telescope in more detail.

Observations of LDN\,1111 were made in 13 hours over two days in 2007 November. Data reduction was performed using the local software tool
\textsc{reduce}. This applies
both automatic and manual flags for interference and
shadowing and hardware errors. It also applies phase and amplitude
calibrations; it then
Fourier transforms the correlator data to synthesize the frequency
channels 
before output to disk in uv FITS format suitable for imaging in
\textsc{aips}.

Flux calibration was performed using short observations of 3C286 near
the beginning and end of each run. We assumed I+Q flux densities for this source in the
AMI SA channels consistent with Baars et al. (1997) $\simeq 3.3$\,Jy at
16\,GHz. As \citet{1977A+A....61...99B} measure I and AMI SA measures
I+Q, these flux densities 
include corrections for the polarisation of the source derived
by interpolating from VLA 5, 8 and 22\,GHz observations. A correction is
also made for the changing intervening air mass over the observation. From
other measurements, we find the flux calibration is accurate to better than
5 per cent (Scaife et~al. 2008; Hurley--Walker et~al., in press).

The phase was calibrated using hourly interleaved observations of the point
source J2201+508 ($\alpha = 22^h 01^m 43^s.5,  \delta = 50^{\circ} 48'
56''.4$), which has a flux density of 0.3\,Jy at 16\,GHz. It was
selected from the Jodrell Bank VLA Survey (JVAS;
\citealt{1992MNRAS.254..655P}). After calibration, the phase is generally stable to
$5^{\circ}$ for channels 4--7, and
$10^{\circ}$ for channels 3 and 8. In this work we use only channels
4--7 due to their superior phase stability.

% section to go in the caption of the figure, if you want it
 The FWHM of the primary beam of the AMI SA is $\approx 20$\arcmin at
16\,GHz. The FWHM of the synthesised beam of the combined channel map towards LDN\,1111 is
$2.4'\times2.1'$ using natural weighting.

\end{subsection}

\begin{subsection}{Imaging}

\begin{figure}
\centerline{\includegraphics[height=8cm,width=9cm,angle=0]{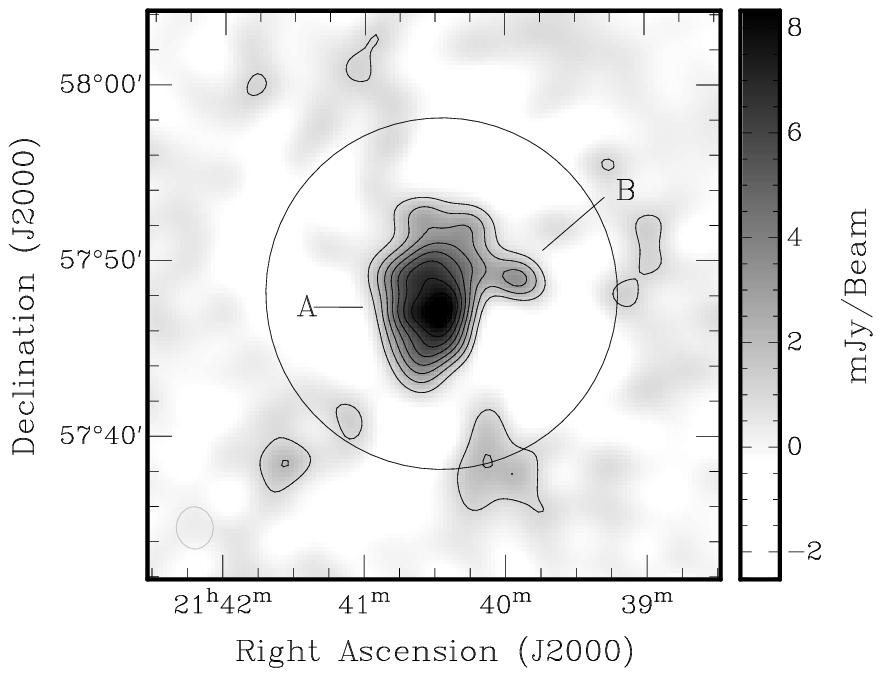}}
\caption{Cleaned combined channel map of LDN\,1111 at 16\,GHz. Contours are increments of
  1\,mJy beam$^{-1}$, with the first contour at 1\,mJy beam$^{-1}$. The primary beam FWHM of the AMI SAis
  shown as a black circle and the synthesized beam as a grey ellipse
  in the bottom left-hand corner. \label{fig:amiplot}}
\end{figure}

\begin{figure}
\centerline{\includegraphics[height=7.5cm,width=8.5cm,angle=0]{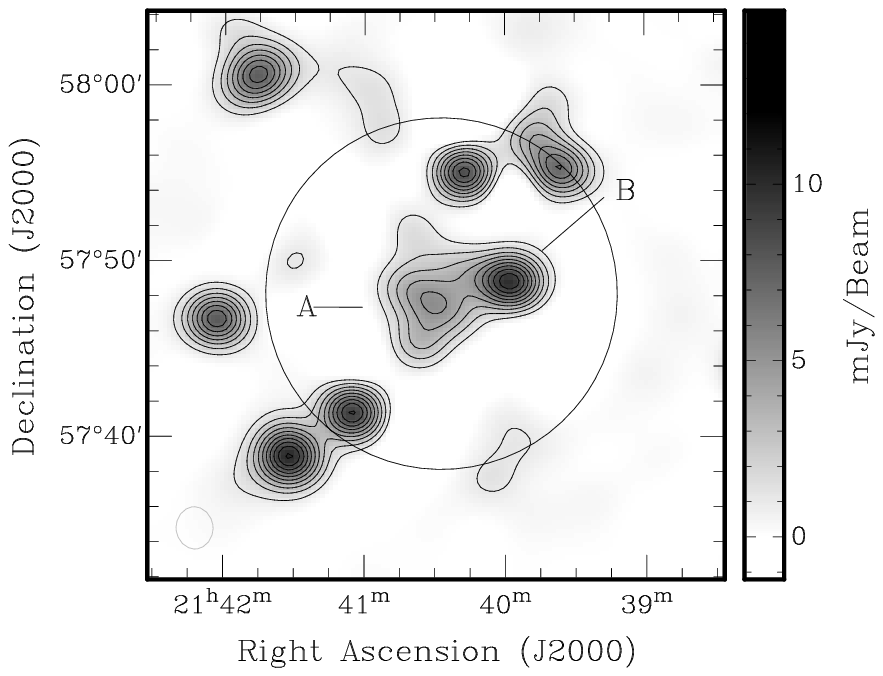}}
\caption{Cleaned map of the $uv$ sampled CGPS data at 1.4\,GHz. Contours are
  increments of 1\,mJy beam$^{-1}$, with the first contour at 1\,mJy beam$^{-1}$. The primary
  beam FWHM of the AMI SA is shown as a black circle and the synthesized beam as a grey ellipse
  in the bottom left hand corner.\label{fig:l1111cgps}}
\end{figure}

The reduced visibilities were imaged using the AIPS data package. Dirty images were
deconvolved using {\sc imagr} which applies a differential
primary beam correction to the {\sc clean} components to account for
the different frequency channels of the AMI instrument. Maps were made
from both the combined channel set, shown here, and for individual
channels. 

Two objects are labelled in both Fig.~\ref{fig:amiplot} and Fig.~\ref{fig:l1111cgps}:
object $A$, which is the dark cloud; and object $B$, which is the NVSS
radio source J213955+574859.

\end{subsection}

\begin{subsection}{Radio Spectra}

\begin{table}
\centering
\caption{Flux densities for  J213955+574859\vspace{0.2cm}\label{tab:l1111src}} 
\begin{tabular}{ccc}
\hline\hline
Frequency & Flux density & Reference\\
(GHz) & (mJy) & \\
\hline
0.327 & 20.0$\pm$4.8 & WENSS; Rengelink et~al. 1997\\
0.408 & 13.4$\pm$1.3 & this work; data from \\
&& Taylor et~al. 2003\\
1.420 & 12.2$\pm$1.25& this work; data from \\
&& Taylor et~al. 2003\\
1.420 & 11.1$\pm$0.5 & NVSS; Condon et~al. 1996\\
14.631& 3.48$\pm$0.56& this work\\
15.381& 2.88$\pm$0.60& this work\\
16.130& 3.02$\pm$0.55& this work\\
17.150& 2.93$\pm$0.57& this work\\
\hline
\end{tabular}
\end{table}

Since the frequency coverage of AMI tells us nothing about the
spectral behaviour at longer radio wavelengths we must combine our new
data with existing archival data. Ideally total power
measurements are required, with the same or better angular resolution than AMI. Data of this type
allow sampling in the $uv$ plane to match exactly the 
measured angular scales of AMI. Here we use data from the Canadian Galactic
Plane Survey (hereinafter CGPS; Taylor et~al. 2003) at
1.42\,GHz. These data are a combination of single-dish and synthesis
data, and
provide a total power measurement of this region with a resolution of
$\approx 1'$. In order to mimic an AMI observation, these data are
modulated by the primary beam of AMI before sampling in the $uv$ plane
to match the AMI $uv$ coverage. These data are 
then processed in the same way as the true AMI visibilities to give
the final sampled image. Fig.~\ref{fig:l1111cgps} shows the sampled
data which result from matching the visibility coverage to AMI
channel 4. 

Spectra of both LDN\,1111 and the radio source
J213955+574859 are presented, identified as objects $A$ and $B$
respectively in Fig.~\ref{fig:amiplot}. Errors on the data points in
these spectra are calculated using a contribution for the rms noise in
each channel, a conservative 5\% flux calibration error and a
contribution for the uncertainty in flux extraction. The morphology of
LDN\,1111 and its surroundings makes this extraction subject to the fitting
area used. To account 
for this in the errors 10 independent `fits' are made. The recorded flux
density is the mean of these measurements and the variance is referred to as
$\sigma^2_{\rm{fit}}$. These errors are combined in quadrature:
$\sigma_{\rm{S}} = \sqrt{\sigma_{\rm{rms}}^2 + (0.05S_{\rm{I}})^2 +
  \sigma^2_{\rm{fit}}}$. The flux extraction is performed using the
{\sc{fitflux}} program (Green 2007), which takes a user defined
polygonal aperture and performs aperture photometry after subtracting
a twisted plane background level. These errors are dominated by the
5\% calibration uncertainty, with the map r.m.s. being $\approx$0.2\,mJy per
channel.

The spectrum of J213955+574859 is presented in order to demonstrate the
flux calibration of AMI, see Fig.~\ref{fig:l1111src}. The data
points for this plot are given in Table~\ref{tab:l1111src}. In
addition to data points from the literature, flux densities at 408\,MHz
and 1.42\,GHz from archival CGPS data are also plotted. These data
have been extracted using 
the {\sc{fitflux}} program as described above. Spectral indices are
calculated using firstly the AMI data 
points plus the literature data, $\alpha=0.52$; and secondly using
all the data 
points, $\alpha=0.46$. These fits are shown as dot-dash and dashed
lines respectively. Both are consistent with $\alpha_{\rm{\sc{wenss}}}^{\rm{\sc{nvss}}}=0.43\pm0.18$. 

For J213955+574859, which is approximately point-like, we expect a
negligible amount of flux loss from extended structure across the AMI bandwidth. In the case
of LDN\,1111 the same is not true. Using the CGPS
map at 1.42\,GHz we find $>$ 50\% flux loss: the integrated
flux from the original unsampled map, calculated as above, is 
$S_{1.42,{\rm{total}}} = 25.4\pm2.7$\,mJy.
However, since the morphology of the source
varies between 1.4 and 16\,GHz we have not used the
CGPS data as a model to calculate flux loss. 

Flux densities for LDN\,1111 are listed in Table~\ref{tab:floss}. It
is immediately obvious that  
the flux densities for LDN\,1111 between 14 and 18\,GHz are
inconsistent with an 
optically thin free--free spectrum extrapolated from 1.42\,GHz. At the
high frequency end of the AMI spectrum there is an excess of almost
50\,mJy, $> 15\sigma_{\rm{S}}\, (\approx 250 \sigma_{\rm{rms}})$. 

Although an absolute measure of the flux loss is not feasible, it is
possible to make an estimate of how the
{\it relative} flux loss will affect the shape of the microwave
spectrum. This is done using a multi-variate Gaussian model for the source with
dimensions of the deconvolved source at 14.63\,GHz (AMI channel 4) as given by the {\sc{aips}} task {\sc{jmfit}}, $7'\times 4'$. This model
is then sampled in $uv$ using the exact 
coverage of the individual channels, and their flux loss relative to
the lowest channel is recorded. We assume an uncertainty of 10\% on the
percentage flux loss and propagate the errors. The results of this calculation are
shown in Fig.~\ref{fig:l1111fullspec} and tabulated in
Table~\ref{tab:floss}. It should be noted that this is a conservative
uncertainty and the relative flux loss will only be in error should
the morphology change significantly between 14 and 17\,GHz.

Fig.~\ref{fig:l1111fullspec} includes a data point at 353\,GHz
(850\,$\mu$m) from the Scuba instrument (Visser, Richer \&
Chandler 2001) for illustartive purposes. This data point is
uncorrected for flux loss. These Scuba data at 850\,$\mu$m are chopped
at 2.5\,arcmin and consequently we 
would assume that information on scales larger than this is lost. To
correctly estimate the flux loss at this frequency we would require
information on the morphology of the vibrational dust emission. Parker
(1988), from whose paper the Scuba sample were selected, lists the
dimensions of LDN 1111 as $1.7 \times 1.1$\,arcmin. A multivariate
Gaussian model based on this data would suggest a flux loss of
approximately 73\%.

Source extraction methods applied to the WMAP data at 23--91\,GHz
(L{\'o}pez-Caniego et~al. 2007) are only able to
give us upper limits on any compact emission towards LDN\,1111, with
the lowest limit being at 41\,GHz, $S_{41} \le 0.30$\,Jy. This value is
consistent with the noise level at that frequency.

\begin{figure}
\centerline{\includegraphics[height=8cm,width=5cm,angle=-90]{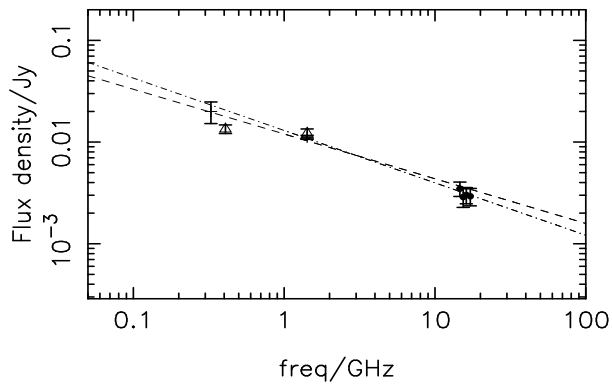}}
\caption{Spectrum of the source J213955+574859. Data points from the
  literature are shown as crosses and are described in the text. Data
  points extracted from the CGPS data are shown as un-filled
  triangles. Data points from the AMI are
  shown as filled circles. \label{fig:l1111src}}
\end{figure}

\begin{table}
\centering
\caption{Flux densities for LDN\,1111 \vspace{0.2cm}\label{tab:floss}} 
\begin{tabular}{cccc}
\hline\hline
Frequency & Flux density & Relative & Corrected \\
& & flux loss & flux density \\
(GHz) & (mJy) & (\%) & (mJy)\\
\hline
1.420$^a$  & 7.2 $\pm$0.7 & -  & 7.2$\pm$0.7  \\
14.631 & 50.7$\pm$2.9 & -  & 50.7$\pm$2.9 \\
15.381 & 53.9$\pm$3.2 & 10 & 59.9 $\pm$6.8\\
16.130 & 54.5$\pm$3.3 & 17 & 65.7 $\pm$7.6\\
17.150 & 55.5$\pm$3.3 & 24 & 73.0 $\pm$8.4\\
\hline
\end{tabular}
\begin{minipage}{7cm}
{\small\vspace{0.1cm} $^{a}$ $uv$ sampled CGPS data point} 
\end{minipage}
\end{table}

\end{subsection}
\end{section}

\begin{section}{Analysis}

The emission seen between 14.63 and 17.15\,GHz by AMI is clearly in
excess of a simple free--free spectrum extrapolated from the CGPS data
at 1.4\,GHz. A number of possibilities may throw some light on the
nature of this excess. The thermal (vibrational) dust spectrum of
proto-planetary disks around T-Tauri stars is expected to extend into
the cm-regime. In these circumstances grain growth has increased the
population of cm-sized grains, or pebbles, causing $\beta$ to approach
zero and the greybody spectrum to fall off with an index approaching
2. However the size of these disks is small and would be unlikely to be
resolved by either AMI or Scuba. LDN\,1111 is quite obviously elongated
in the AMI data, as it is in the Scuba map which has a resolution of
14\,arcsec. 
Although we do not deny the possibility of there being such a disk
embedded within LDN 1111 or projected along the line of sight through
this object, in the absence of further evidence we choose to pursue an
alternative explanation.

A further possibility for a superposition of LDN\,1111 and another
object is that there may be an {\sc Hii} region, with a density such
that it is still optically thick at 16\,GHz, whose flux is contributing
to the spectrum. An {\sc Hii} region with a turn-over frequency
$>30$\,GHz, as would be required here, would possess an emission
measure in excess of $4\times 10^9$\,pc cm$^{-6}$. This would put it
in the regime of ultra-compact and hyper-compact {\sc
  Hii} regions. The spectrum of these objects is inverted with an
index of $\simeq -2$. The AMI data points alone have a spectral index
of $\alpha = -2.86\pm0.35$, rather steeper than would be
expected. Indeed hyper-compact {\sc Hii} regions exhibit slightly
shallower spectra (Franco et~al. 2000) which extends even to
mm-wavelengths. The lack of a visible source in the WMAP data towards
LDN\,1111 would suggest that this is not the case here. In addition
hyper-compact {\sc Hii} regions tend to have very broad radio recombination
line (RRL) widths, typically in excess of 50 km s$^{-1}$ (Kurtz
2005). Ultra-compact {\sc Hii} regions also show broadened line
widths, of typically 30--40 km s$^{-1}$. The RRL measurement of
Heiles, Reach \& Koo (1996) which encompasses LDN\,1111 has a width of
only 21.4 km s$^{-1}$, although we note that this measurement is not
directed at LDN\,1111 specifically. An emission measure of this
magnitude would also imply an {\sc Hii} region mass of around
400\,M$_{\odot}$. The total mass of the cloud from the Scuba data
(Visser et~al. 2002) is 0.3\,M$_{\odot}$ only, which given the
calculated flux loss in this data implies an upper limit on the mass
of 1\,M$_{\odot}$.

A third possibility for the excess emission seen in the AMI data is
that of dipole emission from rapidly rotating small dust grains (Drain
\& Lazarian 1998a,b). The observational evidence for this emission
mechanism has been explored in the introduction to this letter. It is
a relatively new emission mechanism and the evidence for its existence
is not conclusive. We assess the possibility that the emission seen in
LDN\,1111 arises as a consequence of spinning dust by comparing the
data to the model of Draine \& 
Lazarian (1998a;b),which has been used extensively in the past. To
make this comparison we use the MCMC based 
software {\sc metro} (Hobson \& Baldwin 2004) to find the best--fit
parameters.

We fitted a model which has a free--free component normalized to give the
flux density at 1.4\,GHz from the sampled CGPS data with a spectral
index of $\alpha = 0.1$. To this we add a spinning dust component
scaled from the DL98 molecular cloud model.

The model is parameterized by the column
density, $N({\rm{H_2}})$, 
and the angular size of the object. Visser et~al. (2001) calculate the 
average and peak column densities for LDN\,1111 as 5 and 13$\times10^{21}
\rm{cm}^{-2}$, respectively. The results of our model fitting are
$N({\rm{H_2}})$ = 6.72$\pm$0.58$\times 10^{21} {\rm{cm}}^{-2}$, and
$\Delta \theta$ = 5.38$\pm$0.26\,arcmin. These values provide a
$\chi^2$ of 1.03 (79\% probability). However, given the small number of
degrees of freedom in this case we do not propose this statistic as
being conclusive. The resulting model is shown
in Fig.~\ref{fig:l1111fullspec}. The column densities agree well with those of Visser
et~al., although as might be expected there is a degeneracy
between the two parameters.

\begin{figure}

\centerline{\includegraphics[height=8cm,width=5cm,angle=-90]{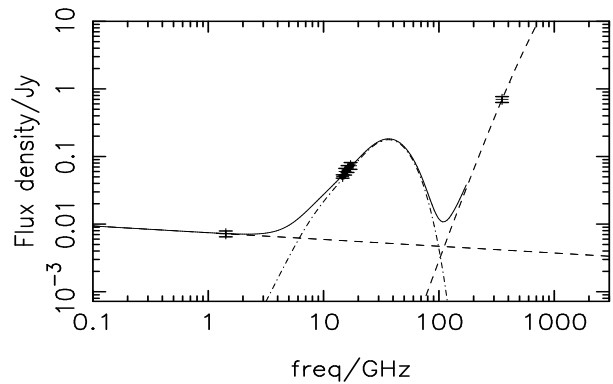}}

\caption{LDN\,1111: Data points are listed in
  Table~\ref{tab:floss}, with the addition of the SCUBA flux
  density at 850\,$\mu$m. A free--free spectrum as described in the
  text and a greybody spectrum scaled to the Scuba data point are
  shown as dashed lines. The best-fit spinning
  dust model of Draine \& Lazarian (1998) is shown as a dot--dash line
  and the total spectrum as a solid line. \label{fig:l1111fullspec}}
\end{figure}

\end{section}

\begin{section}{Discussion and Conclusions}

Given the available evidence we propose the excess of emission we
see towards the dark cloud 
LDN\,1111 in the microwave band to be a result of emission from small
spinning dust grains. This excess is quantified relative to a
flux density derived from lower frequency data sampled to give the
same $uv$ coverage and consequently measuring the same angular
scales on the sky. Lacking further data at lower frequencies with
suitable angular resolution this one point is the basis for the 
assumption that there is an excess in the AMI frequency band. We therefore qualify
this excess by using the original CGPS data to measure the {\it total
  power}, i.e. power on all 
scales, flux density at 
1.42\,GHz to be $S_{1.42,{\rm{total}}} = 25.4\pm2.7$\,mJy. Since
the object is extended this value may be regarded as
an upper limit on the measureable flux at 1.42\,GHz. In addition the total power 
CGPS data at 408\,MHz may then be used to constrain better the lower frequency
spectrum: $S_{408,{\rm{total}}} = 21.6\pm11.8$\,mJy. The large
error on this flux density is dominated by the fitting uncertainty. In
combination these two flux densities yield a spectral index of $\alpha
= -0.16\pm0.12$, which predicts a {\it total power} flux at 16\,GHz of
$S_{16,{\rm{total}}} = 36.9^{+13.8}_{-10.1}$\,mJy. This suggests
that even in terms of total power there is a clear excess at 16\,GHz.

Relative to a canonical free--free spectrum extrapolated
from our sampled CGPS data at 1.42\,GHz we see an excess of
$S_{\rm{excess}} = 60.1\pm7.6$\,mJy, $\approx 8\sigma$, assuming a fixed
spectral index of $\alpha = 0.1$. Using the
spectral index derived from the CGPS total power maps at 408\,MHz and
1.42\,GHz then there is an excess of $S_{\rm{excess}'} = 55.10$\,mJy,
which is still $> 7\sigma$.

In conclusion, we have presented observations of the Lynds' dark nebula LDN\,1111 at
frequencies of 14.6--17.2\,GHz. These measurements show an excess of
emission towards this object relative to an extrapolated free--free
spectrum at a significance of $\approx 15\sigma$. We have proposed that
this excess may be due to emission from small spinning dust grains and
find that it is well-described by the model of Draine
\& Lazarian (1998a;b).

\end{section}

\section{ACKNOWLEDGEMENTS} 

We thank the staff of the Lord's Bridge observatory for their
invaluable assistance in the commissioning and operation of the
Arcminute Microkelvin Imager. We thank Nathalie Ysard and John Richer
for useful 
discussions. We also thank Simon Casassus, whose
comments and suggestions significantly improved this paper. The AMI is supported by Cambridge University and the 
STFC. NHW and MLD  
acknowledge the support of PPARC/STFC studentships.

\bsp
\label{lastpage}

\end{document}